\documentclass[reprint,amsmath,amssymb,aps, prr, nolinenumbers, floatfix,longbibliography,nofootinbib]{revtex4-2}

\usepackage{graphicx}
\usepackage{dcolumn}
\usepackage{bm}
\usepackage[dvipsnames]{xcolor}
\usepackage{physics}
\usepackage{multirow}
\usepackage{booktabs}
\usepackage{comment}
\usepackage{amssymb}
\usepackage{tikz}
\usetikzlibrary{shapes.geometric}
\usepackage{hyperref}
\usepackage{nicematrix}
\NiceMatrixOptions{cell-space-limits = 0.8pt}

\hypersetup{
    unicode=true, 
    plainpages=false,
    colorlinks=true,
    linkcolor=blue,
    citecolor=blue,
    filecolor=blue,
    urlcolor=blue
}

\usepackage{txfonts}

\newcommand{\rbdep}{(\mathbf{r})}

\newcommand{\zeem}{-p\hat{F}_z + q\hat{F}_z^2}
\newcommand{\densint}{\frac{c_0}{2}}
\newcommand{\magnint}{\frac{c_1|\langle\hat{\mathbf{F}}\rangle|^2}{2}}
\newcommand{\spmagn}{|\langle\hat{\mathbf{F}}\rangle|}
\newcommand{\cspin}{\langle\hat{\mathbf{F}}\rangle}
\newcommand{\FMtwo}{\ensuremath{\mathrm{FM}_2}}

\newcommand{\nematic}{\hat{\mathbf{d}}}
\newcommand{\zetaUN}{\ensuremath{\zeta^{\mathrm{UN}}}}
\newcommand{\xhat}{\ensuremath{\hat{\mathbf{x}}}}
\newcommand{\yhat}{\ensuremath{\hat{\mathbf{y}}}}
\newcommand{\zhat}{\ensuremath{\hat{\mathbf{z}}}}
\newcommand{\rhat}{\ensuremath{\hat{\mathbf{r}}}}
\newcommand{\Arg}{\ensuremath{\mathrm{Arg}}}

\begin{document}

\preprint{APS/123-QED}

\title{Composite cores of monopoles and Alice rings in spin-2 Bose-Einstein condensates}

\author{Giuseppe Baio}
\affiliation{Physics, Faculty of Science, and Centre for Photonics and Quantum Science, University of East Anglia, Norwich NR4 7TJ, United Kingdom}

\author{Magnus O. Borgh}
\affiliation{Physics, Faculty of Science, and Centre for Photonics and Quantum Science, University of East Anglia, Norwich NR4 7TJ, United Kingdom}

\date{\today}

\begin{abstract}
We show that energy relaxation causes a point defect in the uniaxial-nematic phase of a spin-2 Bose-Einstein condensate to deform into a spin-Alice ring that exhibits a composite core structure with distinct topology at short and long distances from the singular line. An outer biaxial-nematic core exhibits a spin half-quantum vortex structure with a uniaxial-nematic inner core. By numerical simulation we demonstrate a dynamical oscillation between the spin-Alice ring and a split-core hedgehog configuration via the appearance of ferromagnetic rings with associated vorticity inside an extended core region. We further show that a similar dynamics is exhibited by a spin-Alice ring surrounding a spin-vortex line resulting from the relaxation of a monopole situated on a spin-vortex line in the biaxial-nematic phase. In the cyclic phase similar states are shown instead to form extended phase-mixing cores containing rings with fractional mass circulation or cores whose spatial shape reflect the order-parameter symmetry of cyclic inner core, depending on the initial configuration.
\end{abstract}

\maketitle

\section{Introduction}
Monopoles are point-like topological excitations emerging in seemingly distant physical systems across a vast range of energy scales. They have attracted considerable attention since Dirac's first construction of a magnetic monopole~\cite{Dirac_1931}, where the point charge appears at the termination point of a nodal line. Similar states also appear in superconductors~\cite{Volovik_2000} and superfluids~\cite{Volovik_2003,Savage_2003b,Pietila_2009,Ray_2014}. Monopoles as isolated point singularities are found in contexts ranging from proposed magnetic monopoles~\cite{Hooft_1974, Polyakov_1974} in grand unified theories~\cite{Kibble_1976, Preskill_1979, Weinberg_2007}, to their analogues in condensed-matter systems such as superfluid liquid $^3$He~\cite{Volovik_2003}, chiral superconductors~\cite{Volovik_2000}, and liquid crystals~\cite{Schopohl_1988, Mori_1988, Chuang_1991}, as well as in atomic quantum gases~\cite{Stoof_2001,Ruostekoski_2003,Ray_2015}. 

Despite the very different physical contexts, topologically stable monopoles nevertheless exhibit universal fundamental properties that arise generically from underlying symmetries~\cite{Mermin_1979,Annala_2022}. Their point-like cores are not generally stable, but deform in similar ways into closed line defects: in the context of non-Abelian gauge theories, the core of a 't~Hooft--Polyakov monopole~\cite{Hooft_1974, Polyakov_1974} is predicted to continuously relax into a ring-shaped line defect referred to as an Alice ring~\cite{Bucher_1992, Bais_2002}. 
Observed far from the core, the configuration still preserves the original monopole charge in the form of a ``Cheshire charge'' which does not have a localized source, neither at any particular point nor on the Alice ring~\cite{Alford_1990}. For the same topological reasons, the corresponding point defects in liquid crystals, and atomic quantum gases can similarly deform into Alice rings in the form of disclination loops~\cite{Schopohl_1988, Mori_1988} and closed half-quantum vortices (HQVs)~\cite{Ruostekoski_2003}, respectively. 
The Alice ring (closed Alice string) is universally defined by exhibiting a $\pi$-winding of a suitably defined order parameter in such a way that  a monopole transforms into an anti-monopole after one revolution around a loop through the ring~\cite{Schwarz_1982a,Schwarz_1982b,Bucher_1992,Volovik_1977,Annala_2022}. 

While magnetic monopoles in their original contexts remain elusive~\cite{Milton_2006}, analogues of both Dirac and 't~Hooft--Polyakov monopoles have recently been experimentally realized~\cite{Ray_2014,Ray_2015} in spinor Bose-Einstein condensates (BECs)~\cite{Kawaguchi_2012,Stamper-kurn_2013}. These are created in all-optical traps that avoid freezing out the spin degree of freedom of the constituent atoms~\cite{Stamper-kurn_1998}, resulting in a multi-component condensate wave function. Their several magnetic phases exhibit different broken symmetries that provide a rich variety of topological excitations~\cite{Kawaguchi_2012, Ueda_2014}, which in addition to monopoles include singular~\cite{Yip_1999,Isoshima_2002, Lovegrove_2012,Leonhardt_2000, Zhou_2003, Semenoff_2007, Huhtamaki_2009, Kobayashi_2009, Lovegrove_2012, Borgh_2016b} and non-singular~\cite{Mizushima_2002, Martikainen_2002, Mizushima_2004} vortices and 3D Skyrmions~\cite{Al_Khawaja_2001, Ruostekoski_2001,Battye_2002, Savage_2003a,Tiurev_2018}. 
A long series of experiments realizing vortex nucleation in phase transitions~\cite{Sadler_2006}, realization of HQVs~\cite{Seo_2015}, as well as the controlled creation of 2D~\cite{Leanhardt_2003,Choi_2012a,Choi_2012b} and 3D Skyrmions~\cite{Lee_2018}, singular vortices representing both continuous and discrete symmetries~\cite{Weiss_2019,Xiao_2021,Xiao_2022}, and knots~\cite{Hall_2016,Jayaseelan_2024},  have established spinor-BECs as robust platforms to investigate topological defects and related textures. It has recently been theoretically proposed~\cite{Parmee_2022} and experimentally demonstrated~\cite{Jayaseelan_2024} how similar complex topological objects may be optically created in atomic ensembles.
In particular, spinor-BECs have proven fertile ground for the exploration of the properties, dynamics and decay of imprinted monopoles~\cite{Ollikainen_2017,Blinova_2023}. The Alice ring resulting from the deformation of a point defect in the polar phase of a spin-1 BEC~\cite{Ruostekoski_2003} was recently experimentally observed~\cite{Blinova_2023}. This experimental development has been paralleled by vigorous theoretical work, generalising the theory of unknotted closed vortices~\cite{Nakanishi_1988} to a rigorous topological description of monopoles and Alice rings~\cite{Annala_2022}, as well as analyzing the dynamics of imprinted monopole configurations in spin-1 BECs~\cite{Ruokokoski_2011,Tiurev_2016,Tiurev_2019,Mithun_2022,Kivioja_2023}.

Here we show how the relaxation of a point defect in the uniaxial nematic (UN) phase of a spin-2 BEC leads to the formation of a composite-core spin-Alice ring. The expansion of a point defect into an Alice ring happens as a consequence of the ``hairy ball'' theorem~\cite{Ruostekoski_2003} as the core fills with superfluid in a different magnetic phase. We find that the resulting spin-Alice ring develops a biaxial-nematic (BN) outer core, which itself exhibits a spin-HQV structure with a (vortex free) UN inner core. The spin-Alice ring is therefore an example of a composite defect, which exhibits a hierarchy of different broken symmetries at different length scales such that the core of a singular defect itself harbors a nontrivial defect or texture~\cite{Lovegrove_2014,Lovegrove_2016}. By numerically simulating its dynamics, we find a dynamical oscillation of the core structure between the spin-Alice ring and a split-core hedgehog configuration~\cite{Gartland_1999, Mkaddem_2000}. Transitions between these states occur via ferromagnetic (FM) rings carrying mass circulation appearing internally in the extended core, separating out from the spin-Alice ring state, and later re-emerging from the split-core endpoints with opposite circulation, eventually returning to approximately the original configuration.

Monopole configurations are also possible in the BN phase, where they occur similarly to Dirac monopoles with associated line defects. Also these relax to form spin-HQV ring configurations~\cite{Borgh_2016b}.  
Here we consider specifically the case of an initial monopole attached to a spin-vortex line extending away from the monopole in both directions. We find that the monopole texture relaxes into a spin-HQV ring surrounding the spin-vortex line, and again exhibits dynamical oscillations between the spin-HQV ring and a split-core configuration. 
Finally, we similarly consider monopole textures in the cyclic phase.  
In the case where the spin-vortex line extends away from the monopole in both directions, we find a configuration with two separated ferromagnetic rings with associated fractional mass circulation, while a non-axisymmetric core arises when the monopole forms the termination point of one line defect.

This article is organized as follows: In Sec.~\ref{sec: mean_field} we briefly review the mean-field theory of the spin-2 BEC and relevant topological features of its magnetic phases. In Sec.~\ref{sec: un}, we construct the UN monopole solution and analyse its energy relaxation and the dynamics of the resulting state. In Sec.~\ref{sec: bn_c}, we more briefly describe energy relaxation and dynamics resulting from a BN monopole as well as energy relaxation of monopoles in the cyclic phase, before short concluding remarks in Sec.~\ref{sec: conclusions}. 

\section{Mean-field theory and topology \label{sec: mean_field}}
The spin-2 BEC is described in the Gross-Pitaevskii mean-field theory by a five-component wave function $\Psi\rbdep = \sqrt{n\rbdep}\zeta\rbdep$, which we decompose into the atomic density $n\rbdep = \Psi^\dagger\rbdep\Psi\rbdep$ and a normalized spinor $\zeta\rbdep = [\zeta_2\rbdep,\dots,\zeta_{-2}\rbdep]^\text{T}$ satisfying $\zeta^\dagger\rbdep\zeta\rbdep=1$. The mean-field Hamiltonian density is then~\cite{Kawaguchi_2012}
\begin{align}
    \mathcal{H}[\Psi] = \Psi^\dagger\hat{h}_0\Psi  + \left(\densint + \magnint + \frac{c_2|A_{20}|^2}{2}\right)n^2,
    \label{eq: spin2_mfenergy}
\end{align}
where $\hat{h}_0 = \hbar^2\nabla^2/2M + M\omega^2\mathbf{r}^2/2 \zeem$,  with $M$ being the atomic mass, $\omega$ the frequency of an isotropic harmonic trap, and $p, q$ linear and quadratic Zeeman shifts~\cite{Corney_2006}, which we neglect from this point on. 
The interaction strengths $c_{0,1,2}$ are obtained from the two-particle $s$-wave scattering lengths $a_f$ as $c_0 =4\pi\hbar^2(3a_4+4a_2)/7M, c_2 = 4\pi\hbar^2(a_4-a_2)/7M$ and $c_4=4\pi\hbar^2(3a_4-10a_2+7a_0)$, where $f$ is the total angular momentum of the colliding atoms~\cite{Koashi_2000}.
The first two contributions in Eq.~(\ref{eq: spin2_mfenergy}) are spin-independent and spin-dependent contact interactions, where the condensate spin $\cspin = \zeta^\dagger\hat{\mathbf{F}}\zeta$ is found from the vector $\hat{\mathbf{F}}$ of spin-2 matrices. The third term depends on the formation amplitude of spin-singlet pairs, 
\begin{equation}
    A_{20} = \frac{1}{\sqrt5}(\zeta_0^2 - 2\zeta_{-1}\zeta_{1} + 2\zeta_{-2}\zeta_{2}).
    \label{eq: spin_sing_duo}
\end{equation}

\begin{figure}
    \centering
    \includegraphics[width=.78\columnwidth]{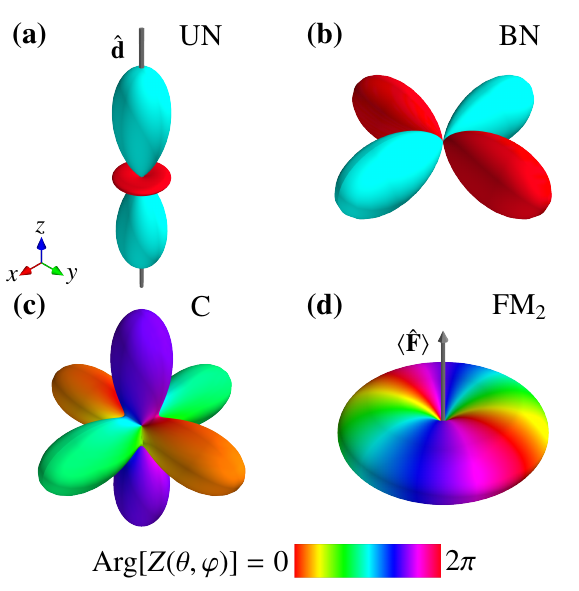}
    \caption{Spherical-harmonics representations [Eq.~\eqref{eq: spherical_harmonics}] for representative spinors of spin-2 magnetic phases. (a) UN phase, whit order parameter given by an unoriented nematic axis $\nematic$ and the condensate phase. (b) BN phase, exhibiting a discrete four-fold dihedral symmetry that combines with the condensate phase. (c) Cyclic phase, with tetrahedral symmetry combined with the condensate phase. (d) $\FMtwo$ phase, with order parameter relating to rotations in three dimensions. The arrow indicates the corresponding condensate spin vector $\langle\mathbf{F}\rangle$.}
    \label{fig: phases}
\end{figure}

In the absence of Zeeman shifts, the spin-2 BEC exhibits four topologically distinct, uniform ground-state phases. Two energetically degenerate nematic phases with $\spmagn=0$ and $|A_{20}|=\sqrt{1/5}$ form ground states in a uniform system when $c_2<0$, $c_2<20 c_1$. The first of these is the UN phase, which can be represented by the spinor
\begin{equation}
    \zeta^\text{UN} = \left(0,0,1,0,0\right)^\text{T}.
    \label{eq: UN_repr}
\end{equation}
The symmetry the order parameter can be illustrated using the spherical-harmonics representation
\begin{equation}
    Z(\theta,\varphi) = \sum_{m=-2}^{2} Y_{2,m}(\theta,\varphi) \zeta_m,
    \label{eq: spherical_harmonics}
\end{equation}
where $\theta$ and $\varphi$ are local spherical coordinates. Applying this representation to Eq.~\eqref{eq: UN_repr} yields Fig.~\ref{fig: phases}(a), showing the symmetry axis $\nematic$. Any other UN spinor can be constructed by applying a shift of the condensate phase (gauge) $\tau$ [an element of $\text{U}(1)$] and a three-dimensional spin rotation [an element of $\text{SO}(3)$], parametrized by three Euler angles $\alpha, \beta, \gamma$ as
\begin{align}
    \zeta \rightarrow e^{i\tau} \exp(-i\alpha\hat{F}_z)\exp(-i\beta\hat{F}_z)\exp(-i\gamma\hat{F}_z) \zeta.
    \label{eq: spin_rot}
\end{align}
This transformation represents the full symmetry group $G=\text{U}(1)\times\text{SO}(3)$ of Eq.~\eqref{eq: spin2_mfenergy}. The isotropy group $H$ of elements of $G$ that leave the spinor unchanged the yields the order parameter manifold as $\mathcal{M}=G/H$~\cite{Mermin_1979}. 
Specifically, it follows from Fig.~\ref{fig: phases}(a) that $\zetaUN$ is unchanged by any rotation about the nematic axis $\nematic$, and also that so does taking $\nematic \to -\nematic$ (nematic symmetry). As a result the UN order-parameter space is $\mathcal{M}^\text{UN} = U(1) \times (S^2/\mathbb{Z}_2)$, where $S^2$ is the 2-sphere~\cite{Zhou_2001}. 

The BN phase, by contrast, is represented by
\begin{equation}
    \zeta^\text{BN} = \frac{1}{\sqrt2}\left(1,0,0,0,1\right)^\text{T},
    \label{eq: BN_repr}
\end{equation}
whose spherical-harmonics representation is shown in Fig.~\ref{fig: phases}(b). The isotropy group is now described by the four-fold dihedral subgroup of $\text{SO}(3)$ combined with $\pi$-shifts of $\tau$, together denoted $(D_4)_{\tau, \mathbf{F}}$, leading to the manifold $\mathcal{M}^\text{BN} =G/(D_4)_{\tau, \mathbf{F}}$~\cite{Borgh_2016b, Kobayashi_2012, Song_2007, Yip_2007}. In addition to the different order-parameter symmetry, the two nematic states can be further distinguished by the amplitude of spin-singlet trio formation
\begin{equation}
    A_{30} \!=\! \frac{3\sqrt{6}}{2}(\zeta_1^2\zeta_{-2} + \zeta_{-1}^2\zeta_2) + \zeta_0(\zeta_0^2 - 3\zeta_1\zeta_{-1} - 6\zeta_2\zeta_{-2}),
    \label{eq: spin_sing_trio}
\end{equation}
such that $|A_{30}|=1$ for the UN and $|A_{30}|=0$ for the BN phases~\cite{Ueda_2002}. 

When $c_1, c_2>0$, the cyclic phase with representative spinor
\begin{equation}
    \zeta^\text{C} = \frac{1}{\sqrt2}\left(1,0,i\sqrt2,0,1\right)^\text{T},
    \label{eq: C_repr}
\end{equation}
whose spherical-harmonics representation is shown in Fig.~\ref{fig: phases}(c), instead becomes the uniform ground state. The interaction contributions now imply $\spmagn=|A_{20}|=0$.  The isotropy group here is described by combinations of elements of the tetrahedral subgroup of $\text{SO}(3)$ with $2\pi/3$-shifts of the condensate phase, together denoted $(T)_{\tau, \mathbf{F}}$.  The order parameter manifold is correspondingly $\mathcal{M}^\text{C}=G/(T)_{\tau, \mathbf{F}}$~\cite{Semenoff_2007, Kobayashi_2012}.

Finally, when $c_1<0$ and $c_2>20 c_1$, energy minimization in the uniform system requires maximizing the condensate spin, yielding a FM phase ($\FMtwo$) with $\spmagn=2$ and $|A_{20}|=0$, represented by 
\begin{align}
    \zeta^{\FMtwo} =\left(1,0,0,0,0\right)^\text{T},  
    \label{eq: FM2_repr} 
\end{align}
whose symmetry is shown in Fig.~\ref{fig: phases}(d) and yields and order-parameter space corresponding to three-dimensional rotations~\cite{Makela_2003,Xiao_2022}.  

Topological defects in ordered media are classified through the homotopy groups of the order-parameter manifold~\cite{Mermin_1979}.
In three dimensions, the first homotopy group $\pi_1(\mathcal{M})$, whose elements are the classes of maps of closed loops into $\mathcal{M}$, identifies singular line defects. Similarly considering maps of closed surfaces into $\mathcal{M}$ yields the second homotopy group $\pi_2(\mathcal{M})$, which characterizes topologically stable point defects. It can also describe topological charges of nonsingular textures whose boundary conditions are fixed, such as 2D Skyrmions.  In the spin-2 BEC, only the UN phase exhibits a non-trivial second homotopy group, $\pi_2(\mathcal{M}^\text{UN})=\mathbb{Z}$. Its integer elements correspond to the monopole charge~\cite{Mermin_1979,Volovik_1977}, 
counting how many times surfaces enclosing the point defects span the manifold $\mathcal{M}^\text{UN}$ in the mapping. For all other spin-2 phases, $\pi_2(\mathcal{M})=0$, i.e., the second homotopy group is trivial and there are consequently no topologically protected, singular point defects. However, they still exhibit solutions corresponding to monopoles attached to one or more singular defect lines~\cite{Borgh_2016b,Baio_2023}, in analogy with Dirac's construction~\cite{Dirac_1931}. In the FM case, this leads to the spin-2 generalisation of the Dirac monopoles in the FM phase of spin-1 BECs~\cite{Savage_2003b}, and can still be characterized by a well-defined winding number as long as the boundary conditions on the spin texture are fixed. This construction generalizes also to the BN~\cite{Borgh_2016b} and cyclic phases in spin-2 BECs.

Relaxation and dynamical stability of the solutions are investigated by simulating the coupled Gross-Pitaevskii equations,  obtained from the energy functional in Eq.~\eqref{eq: spin2_mfenergy},  by means of a split-step method~\cite{Javanainen_2006}, on a Cartesian grid of 128$^3$ points. Interaction strengths ratios are chosen as 
$c_0/c_1=90.7$, $c_0/c_2=-102$, with $Nc_0 = 1.32 \times 10^4 \hbar \omega \ell^3$, corresponding to $N=2\times 10^5$  optically trapped $^{87}$Rb atoms, where $\ell=(\hbar/M\omega)^{1/2}$ and $\omega=2\pi\times 130$ Hz~\cite{Klausen_2001}.  Including Zeeman shifts in Eq.~\eqref{eq: spin2_mfenergy} leads to parameter-dependent ground states~\cite{Koashi_2000, Zhang_2003, Murata_2007}, and restricts the symmetry group $G$~\cite{Kawaguchi_2012}. Here, however, we neglect Zeeman shifts since our primary focus is on the relaxation routes of monopole solutions.  To determine those, we first propagate the solutions in imaginary time. Dynamical stability of the relaxed cores is probed by subsequent complex-time propagation~\cite{Weiss_2019, Xiao_2021, Xiao_2022}.  
 
\section{Uniaxial-nematic monopole and composite-core Alice ring \label{sec: un}} 
An isolated monopole in the UN phase is constructed by taking the nematic axis to coincide with the radius vector everywhere, i.e., $\nematic=\rhat=\cos\varphi \sin\theta\, \xhat +  \sin\varphi \sin\theta\, \yhat + \cos\theta\, \zhat$, where $(\theta, \varphi)$ are spherical coordinates centered on the singular point. Since the state with $\nematic=\zhat$ is exactly Eq.~\eqref{eq: UN_repr}, we can immediately apply Eq.~\eqref{eq: spin_rot} with  Euler angles $\alpha=\varphi$, $\beta=\theta$  and condensate phase $\tau=0$ to find
\begin{equation}
        \zeta^\text{UN}_\text{pm} = \sqrt{\frac38}
        \begin{pmatrix}
                e^{-2i\varphi}\sin^2 \theta \\
               -e^{- i\varphi}\sin 2 \theta \\
                \frac{1}{\sqrt{6}}(1 + 3\cos2\theta) \\
                e^{ i\varphi} \sin 2 \theta \\
                e^{2i\varphi} \sin^2 \theta              
        \end{pmatrix}.   
\label{eq: UN-rotated}    
\end{equation}
 Note that the third Euler angle $\gamma$ drops out and does not need to be specified. 
In exact analogy with the isolated-monopole solution in the spin-1 polar phase~\cite{Ruostekoski_2003} and point defects in superfluid liquid $^3$He and nematic liquid crystals~\cite{Volovik_1977}, a topologically protected charge $W_{\hat{\mathbf{d}}}$ can now be defined as the surface integral
\begin{equation}
    W_{\hat{\mathbf{d}}} = \frac{1}{4\pi}\int_{S^2} d\theta d\varphi \,\hat{\mathbf{d}}\cdot\left(\frac{\partial \hat{\mathbf{d}}}{\partial \theta}\times\frac{\partial \hat{\mathbf{d}}}{\partial \varphi}\right).
    \label{eq: UN_windnum}
\end{equation}
Note that since the nematic director is unoriented the sign of $W_{\nematic}$ is arbitrary, and the charge of the monopole is properly given by $|W_{\nematic}|$.

Isolated-monopole solutions are characterized by an energetically costly core where the superfluid density is depleted and whose size is determined by the density healing length $\xi_n=\ell(\hbar\omega/2nc_0)^{1/2}$~\cite{Stoof_2001}. In spinor BECs, however, defect cores can lower their energy by filling with atoms in a different magnetic phase~\cite{Ruostekoski_2003, Lovegrove_2012}. In a spin-2 BEC, the condensate may then instead break conditions on $\spmagn$ or $|A_{20}|$. The size of the defect core is then instead determined by the spin healing length $\xi_F=\ell(\hbar\omega/2n|c_1|)^{1/2}$ in the former case, or the singlet healing length $\xi_a=\ell(\hbar\omega/2n|c_2|)^{1/2}$ in the latter. Since typically $|c_1|,|c_2| \ll c_0$, and $\xi_F$ and $\xi_a$ correspondingly larger than $\xi_n$, the filled core expands such that the increased interaction energy inside the core is offset by a lower kinetic contribution due to the smaller gradients.  In spin-2, one further possibility arises however: a UN defect singularity can also be accommodated by atoms in the BN phase, or vice versa. The two topologically distinct nematic phases are energetically degenerate at mean-field level. Therefore, unless continuity of the wave function causes mixing with other phases to arise in the core~\cite{Borgh_2016b,Baio_2023} no interaction-energy condition is broken and so the size of the core is in principle not constrained by any healing length (even in the absence of Zeeman level shifts, the degeneracy may, however, still be lifted through order-by-disorder processes beyond mean-field level~\cite{Song_2007,Turner_2007}).

\begin{figure}
    \centering
    \includegraphics[width=\columnwidth]{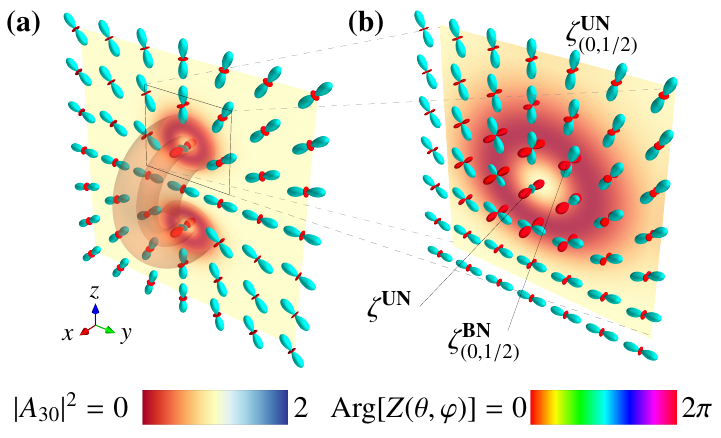}
    \caption{Composite-core spin-Alice ring in the UN phase. (a) Ring defect structure with spin-HQV charge obtained after imaginary time propagation at $\omega t \approx \pi$, shown by $|A_{30}|^2$ in the $yz$-plane (colormap), and isosurfaces at $|A_{30}|^2 = 1/2$. The local order parameter is shown in the $yz$-plane using the spherical-harmonics representation. (b) Detail of the spin-Alice ring core, showing the outer BN core with spin-HQV structure and the inner UN core.}
   \label{fig: UN1}
\end{figure}

For the UN point defect, however, similarly to its analogue in polar spin-1 BECs~\cite{Ruostekoski_2003} and 't~Hooft–Polyakov monopoles~\cite{Bucher_1992, Bais_2002}, filling of its core while maintaining spherical symmetry would imply the existence of a vector field identified by a different local symmetry axis orthogonal to $\nematic$ and thus tangent to $S^2$, which for a point-defect core would violate the "hairy ball" theorem. 
Instead, the singular locus continuously deforms into a closed line defect, shown by numerical energy relaxation in Fig.~\ref{fig: UN1}(a). Here we make use of the spherical-harmonics representation~\eqref{eq: spherical_harmonics} to illustrate the resulting $\nematic$-field. On any surface enclosing the line-defect ring leaves $W_{\hat{\mathbf{d}}}$ unchanged, while $\nematic$ now penetrates the centre of the texture. Along any loop passing through the defect ring, $\nematic$ transforms into $-\nematic$. In contrast to with the spin-1 case, but similarly to nematic liquid crystals, the $\mathbb{Z}_2$ symmetry of the UN order parameter is decoupled from the condensate phase $\tau$, and so the condensate wave function $\Psi$ remains single-valued everywhere without any compensatory phase winding that would lead to mass circulation. The closed line defect resulting from relaxation of the point defect is therefore a spin-HQV, analogous to $\pi$-disclination loops in nematic liquid crystals~\cite{Schopohl_1988,Mori_1988}. Further, the spin-HQV is a spin-Alice ring: in a UN, a monopole of charge $|W_{\nematic}|=1$ turns into its own anti-monopole precisely when encircling a spin-HQV~\cite{Volovik_1977}, establishing the property~\cite{Annala_2022}.

While in the spin-1 case, a singular defect in a polar phase can only develop a ferromagnetic core (and vice versa), the much larger family of stationary states in spin-2 BECs allows the same singular defect to develop very different filled cores depending on parameters~\cite{Borgh_2016b}, experimental choice~\cite{Xiao_2022}, or defect configurations~\cite{Baio_2023}. The resulting core structure may develop considerable complexity as a result of interplay between continuous and polytope order-parameter symmetries. Figure~\ref{fig: UN1} shows how the spin-Alice ring resulting from the relaxation of the UN point defect develops a composite core structure, characterized by a hierarchy of topologically different phases on small and large distances, such that the core of the defect itself exhibits a vortex structure~\cite{Lovegrove_2014, Lovegrove_2016}. Specifically, the order parameter interpolates smoothly between the  the UN and BN phases, from the UN bulk containing the spin-HQV ring, through a BN outer core (characterized by vanishing $|A_{30}|^2$), which itself exhibits a spin-HQV structure [Fig.~\ref{fig: UN1}(b)]. This, in turn, exhibits a (nonrotating) UN core, that forms the inner core of the spin-Alice ring. The topology of the inner and outer cores and the bulk connect by deforming the order parameter along different axes, such that the axis of spin rotation that defines the UN and BN spin-HQVs coincides with one of the symmetry axes of the BN order parameter. This formation of a BN region surrounding a UN ring is closely similar to predictions for a point defect in a nematic liquid crystal confined to a cylindrical capillary tube~\cite{Kralj_1999}.

The transitions within the composite core provide an example of a state with two topological interfaces~\cite{Borgh_2012,Borgh_2014,Lovegrove_2016,Weiss_2019,Xiao_2022}, for which interpolating model wave functions that connect the vortex states can be explicitly constructed~\cite{Baio_2023}. At any point on the vortex spin-HQV line we may defined local cylindrical coordinates and construct the phase-mixing spinor 
\begin{equation}
    \zeta^\text{UN-BN}_{(0,1/2)} = \frac{1}{\sqrt2}\left[D_+(\rho)\zeta^\text{UN}_{(0,1/2)} - D_-(\rho)\zeta^\text{BN}_{(0,1/2)}\right], \label{eq: UNBN_sHQV}
\end{equation}
where
\begin{align}
    \zeta^\text{UN}_{(0,1/2)} &= \frac{1}{\sqrt8}\left(\sqrt3 e^{-i\varphi}, 0, -\sqrt2,0, \sqrt3 e^{i\varphi}\right)^\text{T}, \label{eq: UN_sHQV}\\
    \zeta^\text{BN}_{(0,1/2)} &= \frac{1}{\sqrt8}\left(e^{-i\varphi}, 0, \sqrt6,0, e^{i\varphi}\right)^\text{T}, \label{eq: BN_sHQV}
\end{align}
and $D_\pm(\rho)=[1\pm f(\rho)]^{1/2}$ for a suitable interpolating function $f(\rho)$ of the radial distance. Equations~\eqref{eq: UN_sHQV} and \eqref{eq: BN_sHQV} are obtained, respectively, from Eqs.~\eqref{eq: UN_repr} and \eqref{eq: BN_repr} under the transformation~\eqref{eq: spin_rot} with $\tau=0$, $\alpha=\varphi/2$, and $\beta=\pi/2$. The result in each case is a spin-HQV with no mass circulation which we denote by the subscript $(0,1/2)$. Expressions on the form of Eq.~\eqref{eq: UNBN_sHQV} are very general and can describe several different UN-BN topological-interface structures~\cite{Baio_2023}, including filled vortex cores. For Eq.~\eqref{eq: UNBN_sHQV} specifically, taking $f(\rho)\rightarrow \pm 1$ yield, respectively, the UN and BN spin-HQVs. Additionally, for $f(\rho) = \pm 1/2$, we recover a further BN spin-HQV ($+$) and the vortex-free UN ($-$) spinor. The composite core structure depicted in Fig.~\ref{fig: UN1}(b) can thus be parametrized by taking $f(\rho)$ to vary smoothly from $f(0)=-1/2$ on the vortex line (inner UN core), via $f(\rho^*)=1/2$ in the BN outer core at some distance $\rho^*$ from the singular line, to $f(\rho)\rightarrow 1$ for the UN spin-HQV outside the core region. Since the interaction energy is unchanged between the UN and BN phases, the size of the Alice ring can expand very rapidly during energy relaxation.

\begin{figure}
    \centering
    \includegraphics[width=\columnwidth]{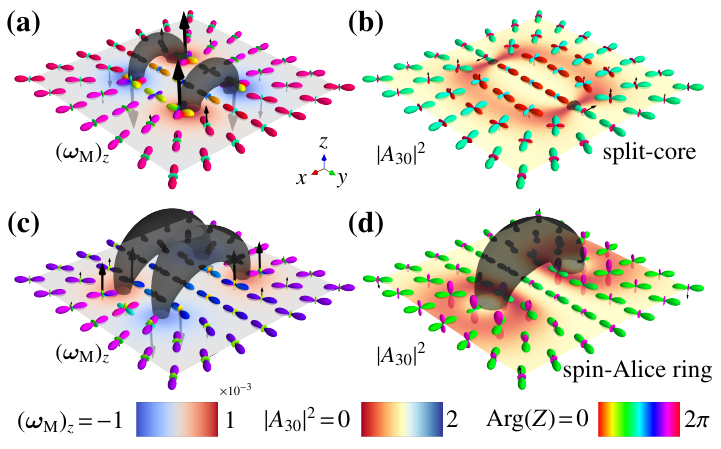}
    \caption{Core oscillations of a spin-Alice ring, shown by frames of complex time simulation. (a) Separation of two vortex rings with FM core at $\omega t\approx 3.2$. (b) Vortex rings collapse into the split-core configuration at $\omega t\approx 7.5$. (c) Vortex rings re-emergence with opposite mass circulation at $\omega t\approx 10.5$ (d) Retrieval of the spin-Alice ring (single ring with maximal spin-vorticity) at $\omega t\approx 15$.
    All panels show the local $\langle \hat{\mathbf{F}}\rangle$ (black arrows), spherical-harmonics representation of the order parameter, and isosurfaces of the spin summed pseudovorticity magnitude. In (a,c), mass circulation in is indicated by the $z$-component of the mass pseudovorticity (colormap). In (b,d) the colormap indicates $|A_{30}|^2$, showing the two nematic core structures of the split-core and spin-Alice ring states, respectively.}
   \label{fig: UN2}
\end{figure}

In order to study its dynamics, we now numerically propagate the spin-Alice ring state in Fig.~\ref{fig: UN1} forward in time, which we give a small imaginary part to phenomenologically account for dissipation. We find that the spin-Alice-ring core exhibits a dynamical oscillation show in Fig.~\ref{fig: UN2}, characterized by the appearance of a pair of FM rings, originating from the spin-Alice ring, with an associated mass circulation. 
The vorticity structure can be unpacked by introducing two pseudovorticity vector fields~\cite{Villois_2016, Wheeler_2021} measuring the local circulation of the mass and spin supercurrents and defined as~\cite{Kawaguchi_2012, Ueda_2014}
\begin{align}
&\boldsymbol{\omega}_\text{M} = -\frac{i\hbar}{2M}\nabla \times \sum^{2}_{\mu=-2} \left(\psi^*_\mu\nabla\psi_\mu - \psi_\mu\nabla\psi^*_\mu \right), 
\label{eq: mass_vort} \\
&\boldsymbol{\omega}^\alpha_\text{S} = -\frac{i\hbar}{2M}\nabla \times\sum^{2}_{\mu, \nu=-2}  (\hat{F}_\alpha)_{\mu,\nu}\left(\psi^*_\mu\nabla\psi_{\nu} - \psi_{\nu}\nabla\psi^*_\mu \right), 
\label{eq: spin_vort}    
\end{align}    
where $\alpha={x,y,z}$. We visualize the mass vorticity in a plane orthogonal to the initial spin-Alice ring orientation in Fig.~\ref{fig: UN1}(a). We choose the plane $z=0$, where $\boldsymbol{\omega}_\text{M}$ is approximately aligned with the $z$-axis. Plotting its $z$-component, we identify regions of nonvanishing mass circulation surrounding the points where $\spmagn$ becomes appreciable. Moreover, the rings exhibit spin circulation, which we visualize by defining the scalar quantity $\sum_\alpha |\boldsymbol{\omega}^\alpha_\text{S}|^2$ (i.e., the sum of the magnitude squared of the $x,y,z$ spin pseudovorticities) as a measure of the total spin circulation. This is shown as three-dimensional isosurfaces in Fig.~\ref{fig: UN2}(a), indicating maximal spin vorticity in two distinct rings. The configuration corresponds to an extended core region associated with the monopole charge, containing the necessary singularity of the bulk UN order parameter.

Quickly after separation, the radii of the spin-vorticity rings shrink, reducing the the corresponding singularities of the UN order parameter to density-depleted points, joined by a segment where the UN order parameter takes the opposite sign, as shown in Fig.~\ref{fig: UN2}(b).  This configuration corresponds to the split-core hedgehog, introduced as a further relaxation channel for 't~Hooft-Polyakov monopoles in high-energy physics~\cite{Bais_2002}, and point defects in nematic liquid crystals~\cite{Gartland_1999, Mkaddem_2000}. The vortex rings later re-emerge from the split-core endpoints with opposite circulation and drift direction, as shown in Fig.~\ref{fig: UN2}(c), and eventually overlap to once again form the spin-Alice ring as in Fig.~\ref{fig: UN2}(d). The oscillation pattern repeats with a period of $\sim 2\pi \times 2.5 \omega^{-1}$.  By tracing $\Arg(Z)$ in the spherical-harmonics representation of the spinor [Eq.~\eqref{eq: spherical_harmonics}], we determine that the condensate phase $\tau$ winds by $\pi$ on a path around any one vortex ring in Figs.~\ref{fig: UN2}(a,c). Since the condensate spin vanishes outside the FM ring, the mass current depends only on $\tau$~\cite{Kawaguchi_2012}, and so this corresponds to the mass-circulation of an HQV. However, as shown in Figs.~\ref{fig: UN2}(a,c), the vortex rings are embedded within the phase-mixing extended core structure, such that no closed path through one (or both) of the FM rings exhibits a single magnetic phase.  Neither ring can therefore be understood as a well-defined topological defect in its own right.

\section{Structure and dynamics of monopole cores in the biaxial-nematic and cyclic phases \label{sec: bn_c}}

\subsection{Biaxial-nematic monopoles and vortex-ring formation}
Although $\pi_2(\mathcal{M}^\text{BN})=0$,  we can explicitly construct solutions in $\mathcal{M}^\text{BN}$ where a monopole texture is attached to a line defect that extends away in opposite directions from the centre of the monopole. In particular, we can construct a radial hedgehog by aligning one of the symmetry axes of the BN order parameter (cf.\ Fig.~\ref{fig: phases}) with the radius vector~\cite{Borgh_2016b}. 
Here we choose one of the two-fold principal axes: Starting from the spinor
\begin{equation}
    \zeta^\text{BN} = \frac{1}{\sqrt8}\left(1,0,\sqrt6,0,1\right)^\text{T}
    \label{eq: BN_repr2}
\end{equation} 
[related to Eq.~\eqref{eq: BN_repr} by a $\beta=\pi/2$ rotation], we apply Eq.~\eqref{eq: spin_rot} with $\alpha=\varphi$ and $\beta=\theta$ ($\tau=\gamma=0$) to form the hedgehog. 
The resulting state is similar to that introduced in liquid crystals to describe the BN ordering around an anchoring colloid~\cite{Alexander_2012} and is described by the spinor
\begin{equation}
        \zeta^\text{BN}_\text{1-pm-1} = \frac{1}{\sqrt8}
    \begin{pmatrix}
        e^{-2i\varphi}\left(1+\sin^2\theta\right)\\
       -e^{-i\varphi} \sin 2\theta\\
        \sqrt{6}\cos^2\theta\\
        e^{ i\varphi} \sin 2\theta\\
        e^{2i\varphi}\left(1+\sin^2\theta\right)      
    \end{pmatrix}.
    \label{eq: BN-rotated} 
\end{equation}

\begin{figure}
    \centering
    \includegraphics[width=\columnwidth]{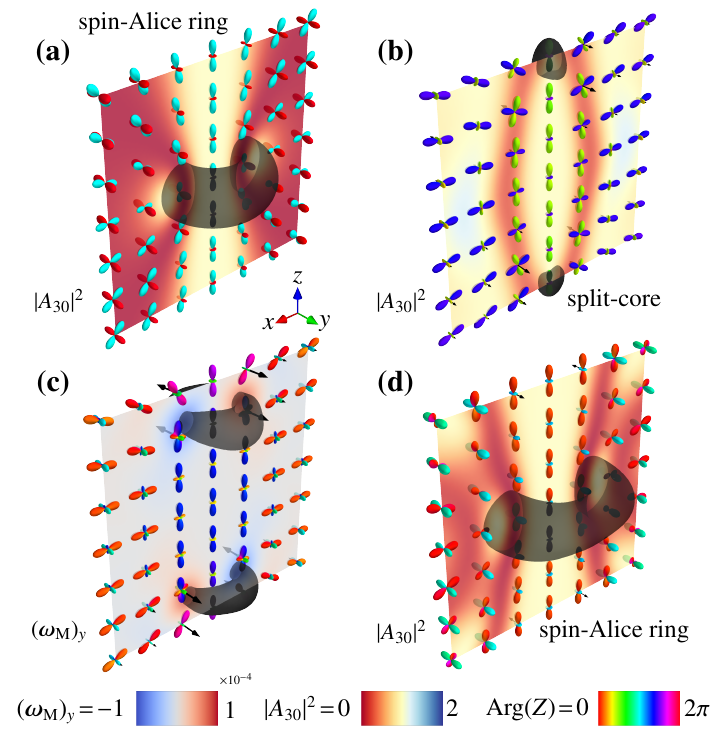}
    \caption{Core oscillations of a BN spin-Alice ring shown by frames of complex time simulation. (a) Core structure of the BN spin-Alice ring penetrated by spin vortex along $z$ as initial state. (b) Split-core-like configuration at $\omega t\approx 12.5$. (c) Vortex rings with mass circulation emerging from the split-core endpoints at $\omega t\approx 14.5$. (d) Retrieval of a spin-Alice ring (single ring with spin-vorticity and no mass circulation) at $\omega t \approx 22$.
    The order parameter is shown by spherical-harmonics representation. The core structures in (a,b,d) are shown by scalar $|A_{30}|^2$ in the $xz$-plane (colormap). In (c) the colormap indicates the $y$-component of mass pseudovorticity.}
   \label{fig: BN}
\end{figure}

Energy relaxation of monopole solutions in the BN phase leads to different core structures depending on the initial configuration of the hedgehog and associated line defects~\cite{Borgh_2016b}. Imaginary-time propagation of the GPEs starting from Eq.~\eqref{eq: BN-rotated} results in the formation of a spin-HQV ring defect, emerging from the hedgehog texture, that encircles the singly-quantized spin-vortex along the $z$-axis. The cores of both defects fill with atoms in the UN phase, as shown in Fig.~\ref{fig: BN}(a). The ring defect exhibits spin circulation , again indicated by an isosurface of $\sum_\alpha |\boldsymbol{\omega}^\alpha_\text{S}|^2$ showing the region of maximal spin-pseudovorticity as defined as in Eq.~\eqref{eq: spin_vort}. The BN spin-HQV ring is similar in structure to the spin-Alice ring in the UN case [comparing Figs.~\ref{fig: UN1} and \ref{fig: BN}(a)], both exhibiting the inversion of an unoriented axis on closed loops around the singular line, without the need for any accompanying shift of the condensate phase. In this sense, the BN spin-HQV may also be thought of as an Alice ring by analogy, despite the absence of point defects in the BN phase. The similarity between the BN spin-HQV ring in the BN phase and the UN spin-Alice ring, is further emphasised by the noting that the core structure of the former may still be locally parametrized using Eq.~\eqref{eq: UN_sHQV}, connecting the spin-HQV-carrying BN bulk order parameter with the UN core across a topological interface. 
In this case, we take $f(\rho)$ to vary from $f(0)=-1/2$ on the line singularity to $f(\rho)=1/2$ for sufficiently large $\rho$ to reach the BN phase away from the vortex line. 

The BN spin-Alice ring also exhibits a closely similar dynamics to its UN counterpart. Complex-time propagation taking the BN spin-Alice ring plus spin vortex configuration in Fig.~\ref{fig: BN}(a) as the initial state also results in oscillations of the core structure analogous to Fig.~\ref{fig: UN2}. Here these are characterized by a first rapid expansion of the UN core in both the spin-Alice ring and the spin vortex along the $z$-axis, giving rise to separating vortex rings with non-vanishing mass- and spin-circulation, drifting away from the centre of the condensate. Also similarly to the UN case (Fig.~\ref{fig: UN2}), the vortex rings shrink to points of density depletion in a split-core-like configuration [Fig.~\ref{fig: BN}(b)], later re-emerging with opposite circulation and inwards drift direction [Fig.~\ref{fig: BN}(c)].  The core structure eventually returns to its original configuration, where the rings overlap to form a BN spin-Alice ring, as shown in Fig.~\ref{fig: BN}(d). 
Despite the similar dynamics, however, the BN configuration deteriorates more rapidly with dissipation and after a longer period of complex-time propagation, the spin-Alice ring is eventually lost.

\subsection{Core structure of a cyclic monopole}
We can apply a construction similar to that leading to Eq.~\eqref{eq: BN-rotated} to construct a monopole state in the cyclic phase. The second homotopy group $\pi_2(\mathcal{M}^{\mathrm{C}})$ is a again trivial and any monopole must come with associated singular lines. Starting from the representative cyclic spinor in Eq.~\eqref{eq: C_repr}, and applying Eq.~\eqref{eq: spin_rot} with Euler angles $\alpha=\varphi$ and $\beta=\theta$ ($\tau=\gamma=0$), we obtain
\begin{equation}
        \zeta^\text{C}_\text{1-pm-1} = \frac{1}{8}
    \begin{pmatrix}
         e^{-2i\varphi}\left[3+i\sqrt3 + 2e^{-i\pi/3}\cos2\theta\right]\\
    4e^{-i\varphi}e^{-i\pi/3}\sin2\theta \\
        i\sqrt{2}\left(1+3\cos2\theta -2i\sqrt3 \sin^2\theta \right)\\
    -4e^{i\varphi}e^{-i\pi/3}\sin2\theta\\
       e^{2i\varphi}\left[3+i\sqrt3 + 2e^{-i\pi/3}\cos2\theta\right] 
    \end{pmatrix}
    \label{eq: C-rotated} 
\end{equation}
representing a radial hedgehog in one of the three principal symmetry axes of the cyclic order parameter (cf.\ Fig.~\ref{fig: phases} situated on a spin vortex extending along the $z$-axis. 

It is also possible to construct a monopole solution more similar to the FM Dirac monopoles~\cite{Savage_2003b}, where the defect line extends away from the monopole in only one direction. In the BN phase, such a monopole was constructed in Ref.~\cite{Borgh_2016b}. We follow the same method here by adding a spin rotation $\gamma=-\varphi$ in Eq.~\eqref{eq: spin_rot} in the construction of the hedgehog. This has the effect of compensating the effect of the rotations $\alpha=\varphi$ and $\beta=\theta$ in such a way that the cyclic order parameter remains nonsingular everywhere except for a singly-quantized spin vortex for along the negative $z$-axis, terminating as the cyclic monopole at the origin. For notational convenience, the resulting spinor can be decomposed as 
\begin{equation}
    \zeta^\text{C}_\text{1-pm} = \frac{1}{\sqrt2}\left(i\zeta^\text{UN}_\text{pm}+\zeta^\text{BN}_\text{1-pm}\right),
\end{equation}
where $\zeta^\text{UN}_\text{pm}$ is defined in Eq.~\eqref{eq: UN-rotated} and 
\begin{equation}
        \zeta^\text{BN}_\text{1-pm} = \frac{1}{\sqrt8}
    \begin{pmatrix}
         2\left(\cos^4\frac{\theta}{2} + e^{-4i\varphi}\sin^4\frac{\theta}{2}\right)\\
         e^{-3i\varphi}\sin\theta\left[\cos\theta-1 + e^{4i\varphi}(\cos\theta+1) \right]\\
        \sqrt{6}\cos2\varphi\sin^2\theta\\
     -e^{-i\varphi}\sin\theta\left[\cos\theta-1 + e^{4i\varphi}(\cos\theta+1) \right]\\
       2\left(\cos^4\frac{\theta}{2} + e^{4i\varphi}\sin^4\frac{\theta}{2}\right)
    \end{pmatrix}
    \label{eq: C-rotated2} 
\end{equation}
coincides with the analogous construction of a BN monopole as the termination point of a spin vortex.

\begin{figure}
    \centering    
 \includegraphics[width=\columnwidth]{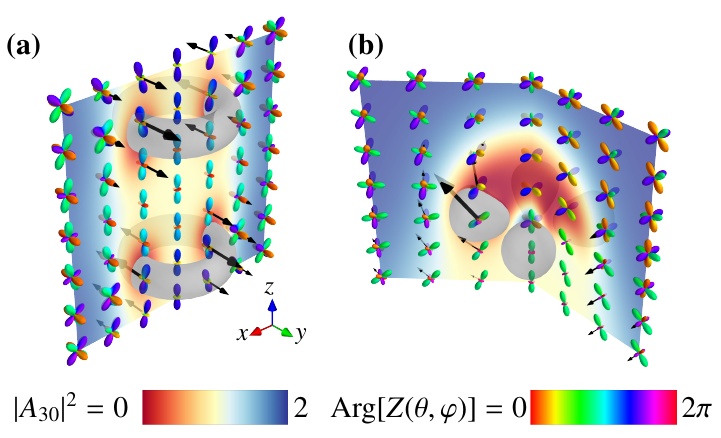}
    \caption{Relaxed core structures of cyclic monopoles, shown by $|A_{30}|^2$ in the $zx$-plane, spherical-harmonics representation of the order parameter, and isosurfaces corresponding to $\spmagn=1$. (a) Cyclic hedgehog with associated spin vortex [Eq.~\eqref{eq: C-rotated}], showing separated vortex rings with FM core. (b) Cyclic monopole as the termination of a spin vortex [Eq.~\eqref{eq: C-rotated2}], showing a non-axisymmetric core structure with localized FM regions.}
   \label{fig: C}
\end{figure}

Considering first energy relaxation of the monopole attached to a spin vortex along the whole $z$-axis, Eq.~\eqref{eq: C-rotated}, we find similarly to the BN case [Fig.~\ref{fig: BN}(a)] that the singly quantized spin vortex develops a UN core extending along $z$. 
By contrast, however, no surrounding spin-vortex ring emerges from the cyclic hedgehog. Instead we find immediate formation of two separated vortex rings with FM core. This is highlighted by the local $\langle\hat{\mathbf{F}}\rangle$ and the isosurfaces with $\spmagn=1$, corresponding to regions with non-zero mass- and spin-vorticity. Each vortex ring displays a fractional $2\pi/3$-winding of the global phase around the FM core, as highlighted from the spherical harmonics in Fig.~\ref{fig: C}(a).  

A very different core structure emerges from energy relaxation of the cyclic monopole when it forms the termination point of a spin-vortex line, given in Eq.~\eqref{eq: C-rotated2}. This  core structure is shown in Fig.~\ref{fig: C}(b), and is breaks the axisymmetry about the $z$-axis in favour of the four-fold symmetry. This reflects the symmetry of the cyclic order parameter as the cyclic phase appears as an inner core at the centre of a now extended, phase-mixing outer core. Specifically this extended core contains four localized FM regions, shown in Fig.~\ref{fig: C}(b) as grey isosurfaces at $\spmagn^2=1$. This reflection of the order-parameter symmetry in the core on the spatial symmetry of the defect core itself is known to appear in the cyclic core of a BN HQV as a result of joining the mismatched point-group symmetries of the core and bulk superfluid~\cite{Borgh_2016b,Kobayashi_2011_preprint}.
In Fig.~\ref{fig: C}(b) we illustrate the relaxed monopole state using two intersecting planes slicing, respectively, through and between FM regions.

\section{Conclusions \label{sec: conclusions}}
In summary, we have demonstrated that a topological point-defect in the UN phase of a spin-2 BEC relaxes to form a spin-Alice ring with a composite-defect core structure consisting of a BN outer core exhibiting a spin-HQV surrounding a UN inner core that accommodates the line singularity. By numerical simulation, we have further shown that the spin-Alice ring exhibits dynamic oscillations between the Alice-ring state and a state corresponding to a split-core monopole configuration analogous to corresponding states in field theories~\cite{Bais_2002} and liquid crystals~\cite{Gartland_1999,Mkaddem_2000}, all while preserving the overall monopole topological (Cheshire) charge. The core oscillations are characterized by the appearance of FM rings with associated mass circulation inside the extended monopole core. In addition, we have also considered monopoles in the BN and cyclic phases, where the second homotopy group is trivial, but states similar to Dirac monopoles, with associated singular line defects, are still possible. We have shown that a BN monopole situated on a spin vortex, which relaxes to a BN spin-Alice ring surrounding the vortex line~\cite{Borgh_2016b}, exhibits a similar oscillatory behaviour between spin-Alice ring and split-core configurations as the true point defect in the UN case. A similar monopole configuration in the cyclic phase, by contrast, does not lead to Alice-ring formation, but we instead find the appearance of an extended core containing FM rings with fractional mass circulation. When the cyclic monopole instead forms the termination point of a spin vortex, energy relaxation leads to the formation of a composite core whose spatial symmetry reflects the order-parameter symmetry of a cyclic-phase inner monopole core. Alice rings have recently been experimentally observed in a spin-1 BEC~\cite{Blinova_2023}. Techniques for engineering of topological defects in all stationary magnetic phases of spin-2 BECs exist~\cite{Xiao_2022}, as well as techniques for creating and observing complex topological objects~\cite{Lee_2018}, which could in principle be used to observe the monopole-core dynamics also in spin-2.

\begin{acknowledgments}
The authors acknowledge financial support from the EPSRC, Grant No.\  EP/V03832X/1. The numerical results presented in this paper were carried out on the High Performance
Computing Cluster supported by the Research and Specialist Computing Support
service at the University of East Anglia.     
\end{acknowledgments}

\bibliography{uea, books}
\end{document}